# Granulation and Convectional Driving on Stellar Surfaces

Johannes Tschernitz[1,2] and Philippe-A. Bourdin[1,3]
[1] University of Graz, Universitätsplatz 5, 8010 Graz, Austria; johannes.tschernitz@tu-braunschweig.de
[2] Now at: TU Braunschweig, Mendelssohnstraße 3, 38106 Braunschweig, Germany
[3] Space Research Institute, Austrian Academy of Sciences, Schmiedlstr. 6, 8042 Graz, Austria

## Abstract

Surface convection is important for the presence of magnetic activity at stars. So far, this convection is thought to be a result of heating from below, where convection cells rise and break up. New models reveal that surface convection is instead strongly driven by cooling from above. We compare two simulations of surface convection, one with a significant heating from below and one without. We obtain surface convection in both cases, and they show similar granulation patterns. The deep convection driven by heating from below is still evolving and asymptotically approaches a steady-state solution. We find that convection from below is not needed at all to form typical photospheric granulation. This indicates the possibility of a surface dynamo acting on stars without a convecting envelope. Even stars without a convecting envelope could therefore exhibit stronger magnetic and coronal activity than expected so far.

*Unified Astronomy Thesaurus concepts:* Solar granulation (1498); Stellar granulation (2102); Solar radiation (1521); Solar convective zone (1998); Stellar convection envelopes (299)

## 1. Introduction

Convection can be described within the framework of mixing-length theory (MLT; E. Böhm-Vitense 1958). Within this framework, the mixing length is the length over which a convecting blob of plasma rises due to buoyancy before dissolving into the background. This length is related to the local pressure scale height. The solar convection zone transports energy within the outermost 29% of the solar radius (J. Christensen-Dalsgaard et al. 1991). There, the temperature and density stratification are unstable, and hence convection occurs (M. Stix 2004).

Convection is observed at different spatial scales throughout the solar interior. On larger scales, there are giant cells (J. G. Beck et al. 1998), supergranulation (A. B. Hart 1956), and there might be mesogranulation (L. J. November et al. 1981; L. J. November et al. 1982) with their respective scales of 100, 20–50, and 5–10 Mm. Å. Nordlund et al. (2009) argue that this distinction is largely historical, and convective motions form a continuous spectrum across all scales rather than being separated phenomena. At the surface of the Sun, the smallest convective cells form the photospheric granules with about 2 Mm diameter (e.g., B. Ruiz Cobo et al. 1996). There, the plasma becomes optically thin and energy is radiated away, leading to a cooling process. The cooled material accumulates and sinks back to the interior in the intergranular lanes (Å. Nordlund et al. 2009).

Since the convection zone is not accessible to direct observations, numerical simulations are widely used to study the properties of convection. Previous simulations were able to reproduce key features of solar convection with realistic temperature and entropy in the convection zone (R. F. Stein & Å. Nordlund 1989, 2000; A. Vögler 2005; B. V. Gudiksen et al. 2011). R. F. Stein & Å. Nordlund (2000) argue that, at least, solar surface convection is driven by radiative cooling in the *photosphere* rather than heating at the base of the convection zone. This would produce low-entropy plasma that forms cool downflows in the darker intergranular lanes (Å. Nordlund et al. 2009), sometimes called "entropy rain" (A. Brandenburg 2016).

Some convection zone models reproduce the general behavior of the largest convective scales by replacing the small-scale photosphere with low-entropy downflows (N. J. Nelson et al. 2018). Later simulations by H. Hotta et al. (2019) show that the inclusion of a proper radiative transfer and photosphere would not significantly impact energy fluxes and rms velocities of the deep convection. The heating from the interior of the star is important in replenishing the lost entropy through radiation but only contributes to the driving of the deep convection on large spatial scales (Å. Nordlund et al. 2009).

The Schwarzschild criterion determines whether a temperature and density stratification in a star is stable against convection. Specifically, a negative entropy gradient is necessary to trigger and drive convection. One study (A. Brandenburg 2016) suggested that MLT should incorporate a nonlocal contribution to the enthalpy flux (J. W. Deardorff 1966). The contribution of this Deardorff flux could allow for a limited mass transport (up and down, but with zero net mass flux) through convection—despite a stable stellar stratification. The horizontal spatial scale of this type of convection is expected to be smaller than regular convection cells.

Simulations demonstrate an amplification of weak magnetic fields near the top of the convection zone (A. Vögler & M. Schüssler 2007; H. Hotta et al. 2015). Surface-driven convection is hence an important mechanism that could amplify small-scale fields and would act like a surface dynamo. Without this additional dynamo action near the surface, the magnetic reconnection in the corona can be expected to be less. Subsequently, the heating in the coronae of stars with no deep convection zone would be higher when we consider surface convection.

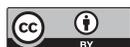

We now set up two 3D hydrodynamic simulations of the upper solar convection zone. The two models are identical except for their treatment of the lower boundary. In one run, the unstable case, we apply an explicit heat input at the bottom of our simulation domain (see Section 2). In the other run, the stable case, we keep the temperature and its gradient fixed at the bottom boundary. This way, we reproduce the unstable solar case and compare it with a stable stratification. We are able to check if the granulation at the surface is ultimately driven from the convection below the surface. Our simulations span a time of about 5 hr of solar time, which is much longer than the typical lifetime of solar granules (about 5 minutes).

## 2. Methods

We use the Pencil Code (The Pencil Code Collaboration et al. 2021) to perform two 3D hydrodynamic simulations of solar convection. We cover a horizontal extent of $64 \times 64$ Mm, and we include the upper 20 Mm of the solar interior up to the photosphere ($\tau = 1$ level). Furthermore, we include 25 Mm of the solar atmosphere above the photosphere in order to allow for a self-consistent evolution of the photosphere. Our grid has $512 \times 512 \times 384$ points, yielding a horizontal resolution of 125 km and a uniform vertical resolution of $\sim 117$ km. We initialize the density and temperature with a solar stratification in hydrostatic equilibrium described in P.-A. Bourdin (2014), truncated to the vertical extent of our simulation box. Throughout the domain, we employ a constant and isotropic heat conduction. The radiative transfer from the photosphere to the atmosphere is modeled by a Newtonian cooling scheme (P.-A. Bourdin 2020), which is smoothly switched on between the photosphere and a height of 1.6 Mm into the atmosphere. The Newtonian cooling scheme pushes the temperature back to the initial stratification according to the local density and with a characteristic half-time of 0.25 s.

The horizontal boundaries are periodic; thus, the stellar curvature is ignored. The top and bottom boundaries are closed for any plasma flows. The top boundary, located in the corona at 25 Mm height, is kept at a fixed temperature that stems from the initial stratification.

We use identical starting conditions for the two runs. The only difference is the treatment of the lower boundary. The first run has a constant heat influx through the bottom boundary. This flux matches the luminosity of the Sun 20 Mm below the surface and is sufficient to drive the deep convection (unstable case). In the second run, the temperature and its gradient at the lower boundary are kept fixed from the initial stratification, which still provides some negligible heat input into the box through conduction. This heat influx is not sufficient to drive the deep convection (stable case). The terms "stable" and "unstable" refer to what is happening deeper in the box in this study.

We solve the following hydrodynamic equations:

$$\frac{D \ln \rho}{Dt} = -\nabla \cdot \boldsymbol{u}, \quad (1)$$

$$\frac{D\boldsymbol{u}}{Dt} = -\frac{1}{\rho} \nabla p - \nabla \Phi_{\mathrm{grav}} + \nu \left( \Delta \boldsymbol{u} + \frac{1}{3} \nabla \nabla \cdot \boldsymbol{u} + 2 \boldsymbol{S} \cdot \nabla \ln \rho \right), \quad (2)$$

$$\frac{D \ln T}{Dt} = -(\gamma - 1) \nabla \cdot \boldsymbol{u} + \frac{1}{\rho c_v T} \nabla \cdot (K \nabla T) + 2 \frac{\nu}{T} \boldsymbol{S}^2, \quad (3)$$

with $\rho$ being the density, $t$ the time, $\boldsymbol{u}$ the velocity, $p$ the pressure, $\Phi_{\mathrm{grav}} = 4\pi G \int_{-R_\odot}^{z'} \rho(z')/z' \, dz'$ the gravitational potential due to the initial density stratification, $\nu$ the kinematic viscosity, $\boldsymbol{S}$ the traceless rate-of-shear tensor, $T$ the temperature, $\gamma$ the adiabatic exponent, $c_v$ the heat capacity for a constant volume, and $K$ the thermal conductivity. The operator $D/Dt \equiv \partial/\partial t + \boldsymbol{u} \cdot \nabla$ is the convective derivative.

As some quantities vary over many orders of magnitude, the continuity (Equation (1)) and energy (Equation (3)) equations are implemented in logarithmic form for an easier numerical treatment due to smaller value ranges.

We use a kinematic viscosity of $\nu = 1.34 \cdot 10^8 \, \mathrm{m}^2 \, \mathrm{s}^{-1}$. This value is just high enough to ensure numerical stability and avoid artifacts in the computed quantities. Significantly lower viscosity would lead to numerical instabilities due to strong velocity gradients at the intergranular lanes. An additional shock viscosity is employed in places with strong shear flows and converging flows that consequently form shocks.

On the initial condition, we impose velocity fluctuations with Gaussian noise in order to break the initial symmetry and hence allow for a more realistic start of the convection. The thermal conductivity is of the same order of magnitude as the kinematic viscosity, giving a Prandtl number around unity. The simulations are purely hydrodynamic and magnetic fields are ignored, which resembles well the situation of fully radiative stars that are expected to have only weak primordial fields if they have not already dissipated through the resistivity of the plasma. We apply a velocity damping to all velocities, which is gradually phased out over the first 6000 s, to allow the initial profile to adapt to the numerical grid and parameter settings without inciting undesired effects (P.-A. Bourdin 2014).

## 3. Results

After 6000 s of solar time, all velocity damping has been faded out smoothly, and the simulation evolves freely. All mentions of time refer to the time passed since the initial condition. After about 2.2 hr the first convection cells become visible in the vertical velocity profile. Regions with significant downflows start to form in the photosphere. In the beginning, downflows are visible only there and extend to a depth of around 4 Mm due to overshooting. At this early point in time, only the upper 1.5 Mm are truly unstable against convection. Since the simulations have not yet reached a relaxed state, all mentions of depths should be taken with caution. The stratification is still evolving, and the depths may change with more simulation time. While the granular cells form, they are relatively small and exhibit a regular pattern. After additional time, their size grows and their shape becomes more irregular. After about 3 hr the granulation has evolved to a state similar to solar granulation, and effects of the initial conditions have vanished. At this time, the photospheric granulation is clearly visible.

Figure 1 shows the temperature gradient $\nabla = d \ln T / d \ln p$ for the two runs after 5.4 hr. In the stable case (dashed–dotted blue line), the horizontal mean gradient in the deep convection zone (below about 2 Mm) remains subadiabatic. This also matches what can be seen in Figure 2(c), where no significant movement is visible below $-5$ Mm. This is different for the unstable case, where we find superadiabaticity below $-17$ Mm. At certain horizontal positions, we find values above 0.4 also up to $-9$ Mm, indicated by the upper red dotted line in Figure 1. A strong influence of the heating from the lower boundary can be seen only in the lowermost 2.5 Mm of the simulation domain. For the upper 2.5 Mm we find convective instability in both cases, which is due to photospheric cooling.

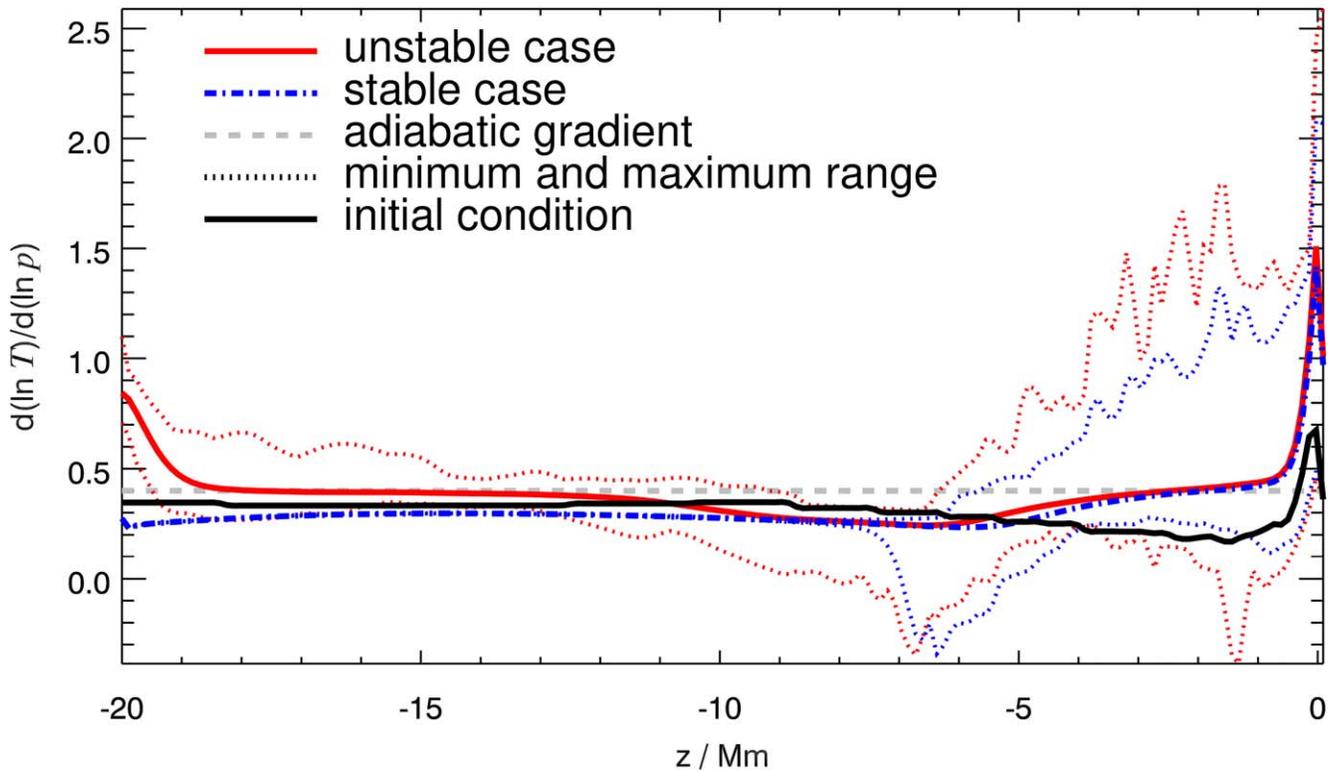

**Figure 1.** Horizontal mean of the temperature gradient, as function of height $z$ from the photosphere, for both cases at 5.4 hr into the simulation. The dotted lines indicate the minimum and maximum of the temperature gradient at each height. The gradient in the stable case is only larger than the adiabatic gradient close to the surface; thus, only this region is unstable against convection. For the unstable case, the mean gradient deeper down is above or close to the adiabatic gradient, indicating unstable conditions. Between $-17$ and $-2.5$ Mm the gradient indicates stable conditions for both runs.

The mean temperature gradients at 10–5 Mm below the surface are similar between the two runs; see the solid red and dashed–dotted blue lines in Figure 1. With the further time evolution of the model, the deep convection and the surface convection still extend in an asymptotic fashion. In the stable case, the expansion of the surface convection ends when there is a balance between low-entropy downflows and thermalization with the ambient plasma. On the real Sun we expect that the undershoot from the deep convection and overshoot from the surface convection may meet and overlap.

With heat influx from the bottom (unstable case), we see large convection cells in the deeper part of our simulation box, which are larger than the granules at the surface; see Figure 2(c). The deep convection cells grow with time and eventually overlap with the surface convection. In contrast, the stable case does not show deep convection, and we see entropy rain from the photosphere down to about 4–5 Mm (see A. Brandenburg 2016); see Figure 2(d).

Already during this early evolution, it becomes clear that surface convection occurs independently from the convection below. The driving mechanism is the radiative cooling in the photosphere.

We now turn to the influence of a deep convection zone on the surface convection. In Figure 2 we show the simulation at an evolved state after 5.4 hr. This is longer than 20 lifetimes of a granule. We show the vertical convective velocities for the unstable case (Figures 2(a) and (c)) and for the stable case (Figures 2(b) and (d)) without the coronal part of the simulation. The photosphere looks qualitatively similar between the two cases with vertical velocities of around a few km s$^{-1}$. The downdrafts exhibit stronger velocities than the upflows due to the mass conservation for large-area upflows with small-area downflows. Smaller granules exhibit larger upward velocities, while in larger granules the upward velocities decrease or even stall in the middle of a larger granule, collapsing into a new downdraft that quickly splits the larger granule into two smaller ones.

The situation below the surface is different. In Figure 2(c), the heat input at the bottom causes the stratification to be unstable against convection, resulting in the formation of convective cells at the bottom of the box separately from the convection at the surface. We see two different zones: The upper $\sim$4 Mm are clearly dominated by the surface-driven convection and overshoot from surface convection. Below $\sim$9 Mm, a deep convection zone has formed. There, vertical velocities are in general smaller, and convection cells have larger diameters, as compared to the surface convection. In principle, this describes two different kinds of convection zones: heating-driven deep convection and cooling-driven surface convection.

The situation in the stable case shown in Figure 2(d) is different: below 2.5 Mm the stratification is stable against convection. Due to no explicit heat input into the system from the bottom, the gradient remains subadiabatic, and no convection is triggered. In both cases, the surface convection cells extend down to about 5 Mm due to overshooting. We find some downdrafts may reach down to 7 Mm (see also Figure 2(c)).

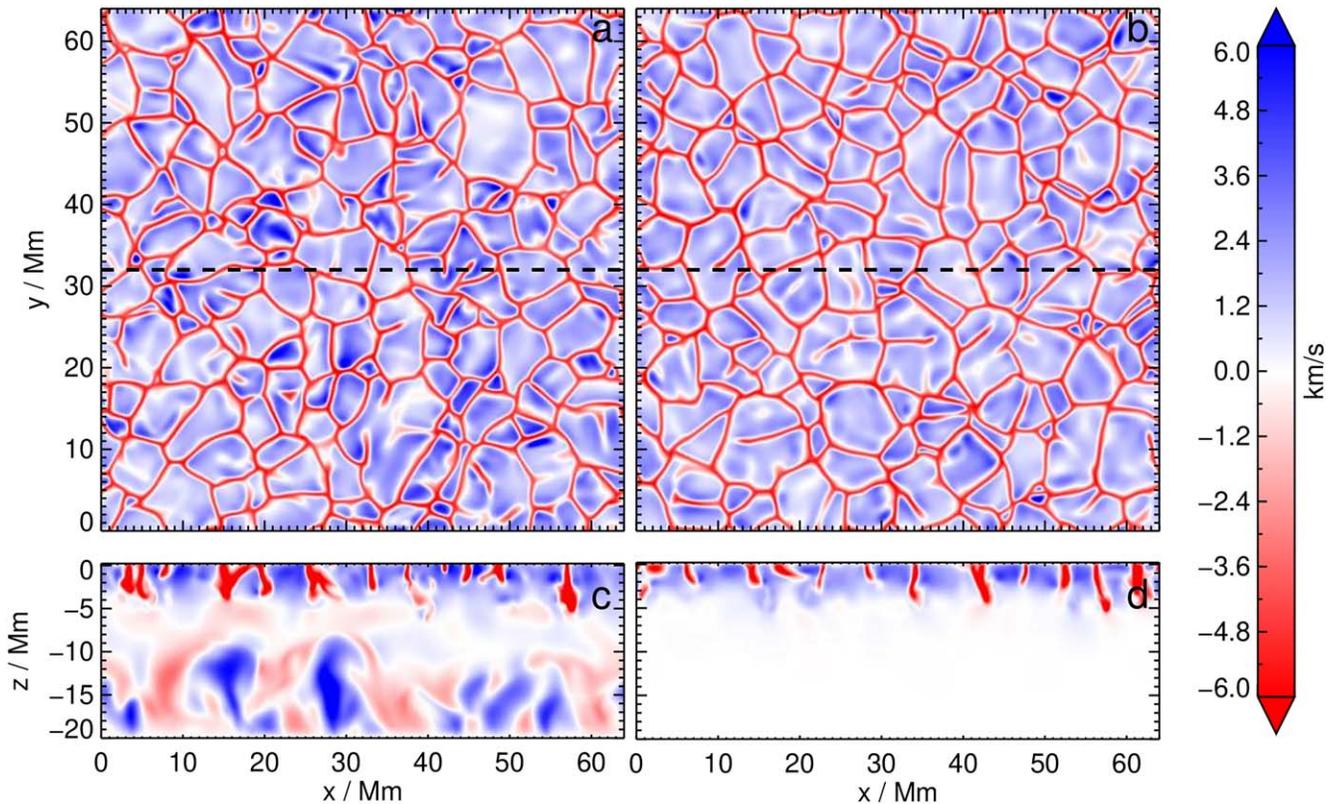

**Figure 2.** (a) and (b) show the vertical velocities in the ($\tau = 1$) layer of our simulation (a) with and (b) without the heating from below after 5.4 hr. (c) and (d) show the vertical velocities in a vertical cut along the black dashed lines in (a) and (b). Saturated blue and red colors correspond to $\pm 6$ km s$^{-1}$.

The sizes and shapes of the granules in the photosphere are very similar between the two runs, with and without a convection zone in the interior; see Figure 2. Also, the vertical velocities are well comparable. Differences in the photosphere between the two runs stem only from random fluctuations in the evolution of the model, since the starting conditions were exactly identical, except for the bottom boundary condition. For the unstable case (Figures 2(a) and (c)) we see that the convection zone is established and reaches the surface convection layer. For the stable case (Figures 2(b) and (d)) we see only the surface convection layer below the photosphere.

Due to the intentional choice of our method, the resulting granules have diameters up to 2.5 times larger than on the real Sun (B. Ruiz Cobo et al. 1996). This is explicable because our viscosity might be too large, and thus we ignore any magnetic fields, and we employ a relatively weak photospheric cooling. The difference to realistic radiative transfer obviously contributes to a larger pressure scale height and hence larger granules. Our model lacks radiative transfer in the convection zone, and we use a single fluid that is fully ionized. However, this does not invalidate our main result that a deep convection zone is not required to trigger surface convection and granulation.

The granular patterns we find in both cases have a similar shape to solar granulation and their intergranular lanes. Vertical velocities are in the expected range, and the intergranular lanes are clearly formed by strong downdrafts.

### 4. Discussion

Both simulations show a solar-like granulation pattern regardless of the heating from below. This demonstrates that photospheric granulation is a pure surface phenomenon and does not require a deep convection zone. In particular, there is no evidence in our simulation that granulation would be a signature of decaying convection cells rising from the convection zone. This finding contradicts common understanding in solar physics. Still, our result agrees with earlier studies (e.g., R. F. Stein & Å. Nordlund 2000; A. Brandenburg 2016).

The stable case has still some heat input from below because we use an isotropic thermal conductivity and the temperature gradient at the bottom boundary is kept fixed. This combination leads to a thermal energy influx into the simulation domain. Still, this heat input is not strong enough to drive the system into instability and trigger deep convection.

Of course, we know from global helioseismology that there is a deep convection zone in the real Sun reaching down to $0.71\,R_\odot$ (e.g., J. Christensen-Dalsgaard et al. 1991; S. Basu & H. M. Antia 1997; J. N. Bahcall et al. 2004). With increasing depth the associated scales get larger and velocities get smaller, although there is some disagreement on how strong these flows are. Studies based on local helioseismology differ by 2 orders of magnitude (S. M. Hanasoge et al. 2012; J. W. Lord et al. 2014; B. J. Greer et al. 2015).

In contrast to common understanding, our results prove that a deep convection zone driven by heating from below is simply not required to produce granulation at the surface of the Sun. This has interesting consequences even for stars without a deep convection zone, in particular for fully radiative stars. We now imply that such stars may still feature shallow, inefficient surface convection, providing a dynamo mechanism that is able to amplify weak or primordial fields to stronger magnetic fluxes than previously expected.

Once we have stronger magnetic fields than expected, there can also be more magnetic activity, such as pores, sunspots, plages, and active regions. This, in combination with the surface-driven granulation, may indeed lead to much stronger coronal heating than previously expected for stars without an outer convection layer. As a consequence, such stars may feature stronger flares and coronal mass ejections. Hence, the subsequently generated energetic particles and high-energy radiation should also be considered regarding the habitability of exoplanets around stars without a deep convection zone.

For a possible confirmation of our result, we suggest to search for signatures of surface convection in exoplanet transits of fully radiative stars, which should reveal an enhancement in light-curve spectra at a characteristic frequency range, which is the inverse of the granule's lifetime. In the case of the Sun, this lifetime is about 3–10 minutes, where an increased power in such spectra occurs. For fully radiative stars, experimental proof of independent surface convection would be to find a Harvey spectrum similar to late-type stars. Alternatively, signatures of convection can be seen in bisectors of spectral lines, which are observationally easier to determine (D. F. Gray & T. Nagel 1989; D. F. Gray 2010).

### Acknowledgments

We thank the anonymous referee for the valuable input. This research was funded in whole or in part by the Austrian Science Fund (FWF) [10.55776/P32958] and [10.55776/P37265]. For open access purposes, the author has applied a CC BY public copyright license to any author-accepted manuscript version arising from this submission.

### ORCID iDs

Johannes Tschernitz  https://orcid.org/0000-0002-2004-222X

Philippe-A. Bourdin  https://orcid.org/0000-0002-6793-601X